# Time-varying Nonlinear Effects in Terahertz Generation

Yongchang Lu[1], Xueqian Zhang[1], Haidi Qiu[1], Li Niu[1], Xieyu Chen[1], Quan Xu[1], Weili Zhang[2], Shuang Zhang[3], Jiaguang Han[1]

[1]Center for Terahertz waves and College of Precision Instrument and Optoelectronics Engineering, Tianjin University and the Key Laboratory of Optoelectronics Information and Technology (Ministry of Education), Tianjin 300072, China.

[2]School of Electrical and Computer Engineering, Oklahoma State University, Stillwater, Oklahoma 74078, USA.

[3]Department of Physics, Faculty of Science, University of Hong Kong, Hong Kang

**Abstract**

Time-varying effects have unveiled new possibilities for manipulating electromagnetic waves through the temporal dimension. In this study, we experimentally explore these effects in the nonlinear optical process of terahertz (THz) generation using optically pumped indium tin oxide (ITO) films. The ultrafast carrier dynamics in the ITO film endow the second-order nonlinear susceptibility ($\chi^{(2)}$) with sub-picosecond temporal evolution, establishing a temporal boundary for the generated THz waves. We observe significant amplitude and frequency modulations in the THz generation at various transients, attributed to the time-varying complex amplitude of the $\chi^{(2)}$. Moreover, we also observed polarization modulations when further exploiting the tensor properties of $\chi^{(2)}$. This work advances the exploration of time-varying effects into the nonlinear regime through frequency down-conversion, effectively transferring the strong time-varying material response from the near-infrared (NIR) band to the THz band. These findings open up new opportunities for realizing time-varying phenomena that specifically require single-cycle modulation.

**Introduction**

Time-varying effects have recently garnered considerable attention, offering new opportunities to exploit the temporal dimension as an additional degree of freedom in the engineering of electromagnetic waves[1-3]. These phenomena describe light-matter interactions in media where the time-reversal symmetry is broken by rapidly changing the permittivity in time. In time-varying systems with spatial symmetry, photon behavior is governed by momentum conservation rather than energy conservation[4]. Owing to the inherent space-time duality in Maxwell's equations, temporal analogues of spatial phenomena—such as temporal boundaries, along with the associated temporal refraction and reflection—can be defined[5,6]. As such, a variety of novel physical effects have been explored, including linear adiabatic frequency shifts[7-9], double-slit time diffraction[10], and momentum gaps[11-13], which emerge at single or dual temporal boundaries, as well as in photonic time crystals (PTCs)[14-16]. Additionally, intriguing concepts such as optical non-reciprocity[17-19], time reversal[20], time aiming[21], and the inverse prism effect[22] have also been proposed, further expanding the scope of time-varying research.

Recently, transparent conducting oxides (TCOs) like indium tin oxide (ITO) and aluminum zinc oxide (AZO) have emerged as novel photonic platforms due to their epsilon-near-zero (ENZ) effects and the tunable optical responses[23,24]. When pumped in the near infrared (NIR) band, TCOs exhibit carrier dynamics over a few hundred femtoseconds dominated by intraband transitions, which enables significant temporal modulation of their linear optical properties[25,26]. Consequently, experimental demonstrations of time-varying effects, such as time refraction[8,9], diffraction[10,27] and linear frequency shift[7,8,28], have been achieved. Additionally, time-varying semiconductors are also exploited to realize linear frequency conversion in the THz regime[29-31], where temporal boundaries are created by rapidly changing the photocarriers concentrations through interband transition. However, the speed of the material response in these experiments is typically much slower than the carrier frequency of the probe pulses. This limitation hinders the realization of certain time-varying applications, such as time reflection and PTCs, which specifically require single-cycle modulation[32].

Nonlinear optics represents a cross-scale light-matter interaction in time and frequency dimension[33]. For instance, the second-order optical rectification (SOR) effect in nonlinear media can generate a new electromagnetic pulse with an oscillation period that matches the pulse width of the incident laser. Notable demonstrations include broadband THz generation in ITO film through the surface SOR effect[34], which is significantly enhanced by the ENZ effect under oblique pumping. As a result, the ultrafast modulation implemented to the linear property of ITO film can be perceived by the nonlinear oscillation for THz emission. This nonlinear process can not only transfer the strong response from the NIR band to the THz band, but also reduce the carrier frequency of the probe pulses through frequency down conversion, allowing for the rate-matching between the material response and the carrier frequency. However, most current research has focused primarily on linear optical processes between the probe pulses and the time-varying media, leaving such nonlinear interactions largely unexplored.

In this work, we experimentally investigated the nonlinear temporal manipulation of THz generation using optically pumped ITO film. The ITO film was pumped at oblique incidence with *p*-polarized NIR femtosecond laser pulses, modulating the THz generation of probe pulses that were separated from the pump pulses. By adjusting the time delay between the pump and probe pulses, we observed a broadband amplitude modulation of 80% and a central frequency modulation of 15% in the generated THz waves. These phenomena can be intuitively understood through a time-varying nonlinear polarization model, where the modulation of the linear susceptibility leads to the ultrafast change in the complex amplitude of $\chi^{(2)}$, thereby determining the temporal evolution of the generated THz waves. Furthermore, when the probe pulses were switched to circular polarization, both the polarization orientation and ellipticity of the generated THz waves exhibited broadband modulation as the time delay changes. This is attributed to the changes in the relative complex amplitudes of the tensor elements of $\chi^{(2)}$. Our experiments demonstrated a significant temporal modulation in the THz band through nonlinear interaction, offering an alternative paradigm for studying time-varying effects.

**Time-varying Nonlinear Polarization**

In a perturbation framework, where the light-matter interaction slightly influences the material property, the SOR-induced THz generation can be described using the typical nonlinear polarization equation:

$$P(\omega_{THz};\omega_1,\omega_2) \propto \chi^{(2)}(\omega_{THz};\omega_1,\omega_2) E(\omega_1) E(\omega_2), \tag{1}$$

here $P$ is the nonlinear polarization; $\chi^{(2)}$ is the second-order nonlinear susceptibility; $E(\omega_1)$ and $E(\omega_2)$ are the electric fields of the incident light with carrier frequencies of $\omega_1$ and $\omega_2$. In the assumption that $\chi^{(2)}$ is time independent, energy conservation associates the generated frequency $\omega_{THz}$ with the fundamental frequencies by $\omega_{THz} = \omega_1 + \omega_2$. Under the perturbation approximation, $\chi^{(2)}$ can be expressed as the product of the linear susceptibilities at three frequencies involved in the nonlinear process:

$$\chi^{(2)}(\omega_{THz};\omega_1,\omega_2) \propto \chi^{(1)}(\omega_{THz}) \chi^{(1)}(\omega_1) \chi^{(1)}(\omega_2), \tag{2}$$

where $\chi^{(1)}$ represents the linear susceptibility. In the context of nonlinear interaction between the nonlinear media and a laser pulse consisting of finite electric field components around carrier

frequency, the total nonlinear polarization at the THz frequency $\omega_{THz}$ can be rewritten as:

$$\begin{aligned}P(\omega_{THz}) &\propto \chi^{(1)}(\omega_{THz})\int_{-\infty}^{+\infty}\chi^{(1)}(\omega_1)\chi^{(1)}(\omega_{THz}-\omega_1)E(\omega_1)E(\omega_{THz}-\omega_1)d\omega_1\\ &=\chi^{(1)}(\omega_{THz})\left[\chi^{(1)}(\omega_{THz})E(\omega_{THz})\right]*\left[\chi^{(1)}(\omega_{THz})E(\omega_{THz})\right]\\ &=\chi^{(1)}(\omega_{THz})\left[\chi^{(1)}(\omega_1)E(\omega_1)\right]*\left[\chi^{(1)}(\omega_1)E(\omega_1)\right]\end{aligned} \quad (3)$$

where the asterisk (*) represents the convolution operation. This equation reveals that the polarization spectrum is proportional to the convolution of $\chi^{(1)}(\omega_1)E(\omega_1)$ with itself. From equation (2) and (3), we can infer that controlling the linear optical property of material allow for direct manipulation of nonlinear THz generation. If the material's properties can be switched at the speed of THz oscillations, it is expected that the THz waveform would undergo temporal modulations. This modulation results from energy coupling between photons and materials, leading to manipulation of the THz spectrum in the frequency domain.

The proposed ultrafast nonlinear dynamics is schematically illustrated in Figure 1, where the significant modification to static nonlinear processes by a pump pulse (represented by the orange envelope) introduces new frequency components into the final THz spectrum. In the dynamical nonlinear framework, the pump pulse acts as an external stimulus to the nonlinear media and is injected at time $t'$, where the time origin coincides with the intensity peak of the probe pulse. This injection rapidly alters the linear susceptibility of material. Due to this material change around $t'$, the series of nonlinear polarization waves oscillating at $\omega_{THz,n}$, which result from the Fourier decomposition of the static convolution of two probe pulses (green envelopes), are either stressed or squeezed. This is equivalent to a change in the complex amplitude of the nonlinear susceptibility. Consequently, a new polarization waveform and its corresponding THz spectrum emerge from the superposition of these modified nonlinear polarization waves.

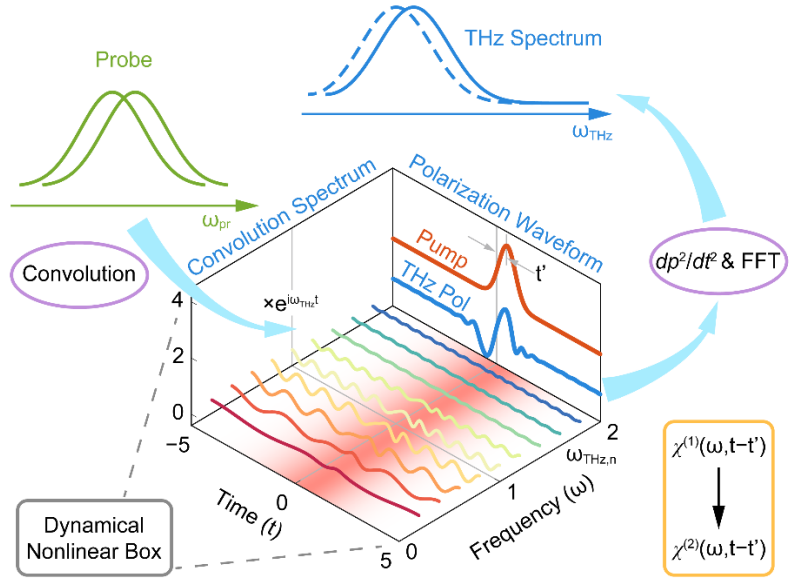

**Figure 1.** Schematic of the dynamical nonlinear polarization model. The original static nonlinear process is empowered temporal dynamics by the pump pulse injected at time $t'$. In the nonlinear dynamical nonlinear box, the pump pulse rapidly changes the material property and then modulates the temporal waveforms of THz polarization, which introduces the fast phase change and consequently the new frequency components.

For an ultrathin ITO film to be investigated here, its relative permittivity can be described using the Drude model:

$$\varepsilon_r = \varepsilon_\infty - \frac{\omega_p^2}{\omega(\omega+i\gamma)}, \tag{4}$$

where $\varepsilon_\infty$ is the infinite frequency permittivity, $\omega_p = \sqrt{ne^2/(\varepsilon_0 m^*)}$ is the static plasma frequency determined by the carrier density $n$ and effective mass of the free electrons $m^*$, and $\gamma$ is the collision frequency. It has been demonstrated that ITO film supports a Brewster mode around the ENZ wavelength[35], where the normal component of the incident fields can be significantly enhanced according to the boundary conditions. Due to the non-parabolic conduction band of ITO[36], a strong pump pulse around ENZ wavelength can significantly modify the permittivity of the ITO film by alerting the effective mass of free electrons through the excitation of intraband transition. This modification introduces a time dependence to the plasma frequency, which ultimately leads to a time-varying permittivity $\varepsilon_r(t - t')$ and thus a linear susceptibility $\chi^{(1)}(t - t') = 1 - \varepsilon_r(t - t')$, where $t'$ is the time delay between the pump pulse and the probe pulse. Therefore, the second-order nonlinear susceptibility also becomes time-varying, as well as the associated nonlinear polarization, which can be described as:

$$\chi^{(2)}(\omega_{THz};\omega_1,\omega_2;t-t') \propto \chi^{(1)}(\omega_{THz};t-t')\chi^{(1)}(\omega_1;t-t')\chi^{(1)}(\omega_2;t-t'), \tag{5}$$

$$P(\omega_{THz};t-t') \propto \chi^{(1)}(\omega_{THz};t-t')\left[\chi^{(1)}(\omega_1;t-t')E(\omega_1;t)\right]*\left[\chi^{(1)}(\omega_1;t-t')E(\omega_1;t)\right]. \tag{6}$$

Clearly, the extent of modulation in the THz spectrum directly depends on the duration of the pump pulse, the material response and the pump-probe time delay $t'$. Based on these equations, numerical simulations were performed to model the THz generation in time-varying media, providing deeper insights into the dynamic behavior of the system (see supplementary materials note 1).

**Experimental Section**

To investigate this nonlinear time-varying system, we used NIR femtosecond laser pulses to pump a 50 nm-thick ITO film, thereby modulating THz generation. The ITO film was deposited on an optical glass substrate via electron beam evaporation and subsequently annealed in a tube furnace (see Methods for details). The static optical constants of the ITO film were characterized using a commercial ellipsometer. As shown in Figure 2a, the gray dashed line indicates the ENZ wavelength around 1563 nm, where the real part (blue line) of the permittivity crosses zero point, and the imaginary part (orange line) is less than unity.

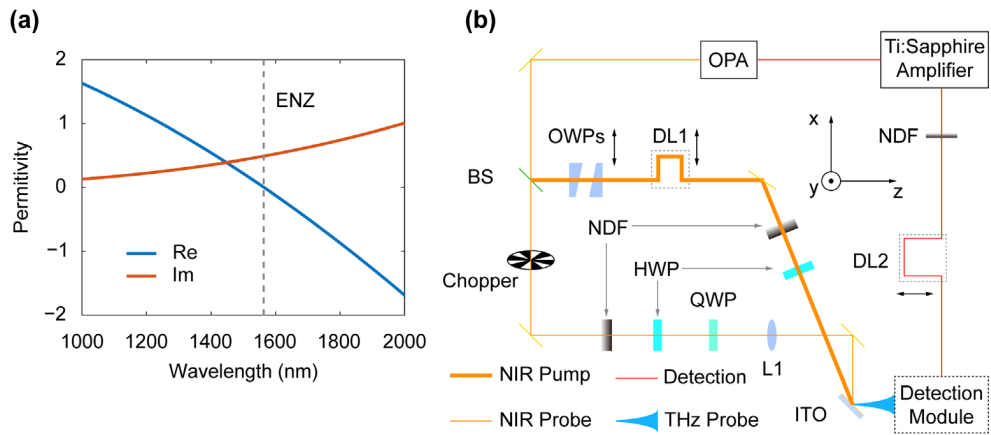

**Figure 2.** Permittivity of the ITO film and schematic of the experimental setup. (a) The real part (blue) and imaginary part (orange) of the fitted permittivity of the ITO film. The gray dashed line indicates the ENZ wavelength where the real part approaches zero. (b) The main components of the experimental setup for optical pumping and nonlinear probing, as well as a brief illustration of the standard THz electro-optic detection. Abbreviations: OPA, optical parameter amplifier; BS, beam splitter; OWPs, optical wedge pairs; DL, delay line; NDF, neutral density filter; HWP, half-wave plate; QWP, quarter-wave plate; L1, lens.

The THz generation was characterized using a homemade THz time-domain spectroscopy (THz-TDS) system. As illustrated in Figure 2b, the setup was powered by a Ti:sapphire femtosecond laser amplifier. The main output energy (3.5 mJ) was used to drive the optical parametric amplifier (OPA) to generate the pump and probe pulses in the NIR band. The pump pulses, with a pulse duration of 45 fs and a central wavelength of 1500 nm, were reflected by a beam splitter (BS) and then passed through a pair of optical wedges (OWPs) and a delay line (DL1), both mounted on motorized translation stages for fine and coarse tuning of the pump beam's optical path. The probe pulses, transmitted through the BS, were modulated by an optical chopper, which provided a reference frequency of 370 Hz to enhance the signal-to-noise ratio during signal acquisition. For both the pump and probe beams, a neutral density filter (NDF) and an achromatic half-wave plate (HWP) were used to control pulse energy and polarization, while a quarter-wave plate (QWP) in the probe beam was used to achieve circular polarization. A lens (L1) with a focal length of 500 mm slightly reduced the probe spot size to ensure optimal overlap with the pump spot. The incident angle of the probe beam was set to 45° relative to the normal of the ITO film, while the pump beam's angle was set to 67° to efficiently excite the Brewster mode, which enhances the modulation of the ITO film (see supplementary materials note 2). The residual energy from the Ti:sapphire femtosecond laser amplifier was attenuated by a NDF and guided to the THz detection module through another delay line (DL2) for standard electro-optic sampling using ZnTe crystal. It worth noting that although both the pump and probe pulses have the capability to generate THz waves, the optical chopper ensured that only THz waves generated by the probe beam were detected. Additionally, the THz waves resulting from the pump-probe interaction in the ITO film had a radiation angle of 79° relative to the optical axis (z-axis), which significantly exceeded the collection range of the detection module.

**Result and Discussion**

In the experiment, maintaining a moderate probe intensity is crucial to facilitate the nonlinear interaction in the ITO film while avoiding significant modulation of the film by the probe pulse itself. The optimal probe intensity was determined to be 7 GW/cm$^2$ by characterizing the dependence of THz amplitude on the probe intensity (see supplementary materials note 3). With this optimal probe intensity and using *p*-polarized pump pulses, we measured the changes in THz electric fields at a given sampling time position as a function of both pump intensity and the time delay between the pump and probe pulses.

As shown in Figure 3a, at a given pump intensity, the peak amplitudes of the THz electric fields undergo asymmetric suppression and recovery dynamics within 1.3 ps time delay. Initially, the pump pulses lag behind the probe pulses and gradually catch up as the time delay increases. The amplitude modulation of the generated THz waves is attributed to the rapidly varying nonlinear susceptibility amplitude, as described in equation (5), while the asymmetric feature is determined by the convolution of the pump envelope and the ITO response. As the pump intensity increases from 0 GW/cm² to 40 GW/cm² in increment of 3.3 GW/cm², the suppression of peak amplitudes monotonically intensifies, approaching saturation beyond a pump intensity of 40 GW/cm². The minima of each curve, marked by purple asterisks, indicate that the time delay corresponding to the strongest modulation shows a slight deviation under higher pump intensities, with the THz amplitude nearly fully suppressed at 43.3 GW/cm². This deviation may be attributed to the modification of the inherent response of hot electrons in the ITO film by the stronger pump, which is linked to the population of electrons in excited states. The nearly vanished THz amplitude is likely an artifact caused by changes in the temporal waveform of the generated THz waves.

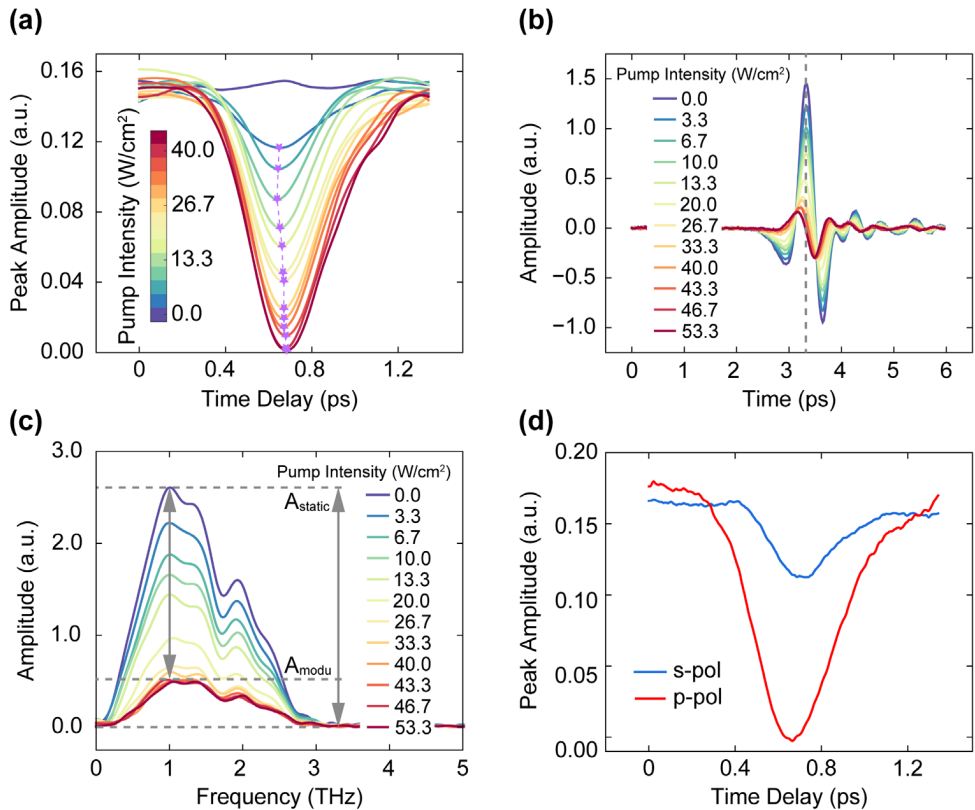

**Figure 3.** Measured THz amplitude modulations versus time delay under various pump intensities with p-polarized pumps and the modulation comparison with s-polarized pumps. (a) Temporal modulated THz electric field under various pump intensities, sampled at a given detection pulse position. The pump intensities were represented using the color bar and the minima in each curve were marked with the purple asterisks. (b) Complete THz electric field traces under various pump intensities were plotted, where the gray dashed line indicated the sampling position in Fig. 3(a). A saturation phenomenon was observed when the pump intensity exceeds 40 GW/cm². (c) Corresponding THz spectra of the temporal signals in Fig. 3(b) were plotted and showed significant broadband amplitude modulations of the generated THz waves, where the modulation depth was defined using the spectral extrema under 0 GW/cm² and 40 GW/cm² pump intensities as indicated by the gray arrows and dashed lines. (d) Comparison of the amplitude modulation effects between the p-polarized and s-polarized pumps. The larger modulation under p-polarized pump was observed and was attributed to the efficient excitation of the Brewster mode.

At the specified pump-probe time delay of 0.68 ps, we measured the complete traces of the THz electric fields at various pump intensities, as shown in Figure 3b, where the gray dashed line indicates the sampling position in Fig. 3a. As the pump intensity increased, the peaks of THz electric fields gradually reduced. However, even at a pump intensity of 53.3 GW/cm², a weak but distinct THz signal was still present, with the gray dashed line crossing over its zero point of the slightly evolving waveform, as seen in Fig. 3a. The corresponding THz spectra, shown in Figure 3c, exhibit a broadband modulation over the generated THz spectrum. The gray dashed lines and arrows indicate the maximal amplitudes $A_{static}$ and $A_{modu}$, measured at 0 GW/cm² and 40 GW/cm² pump intensities, respectively. The amplitude modulation depth, defined as $M_A = \frac{A_{static} - A_{modu}}{A_{static}}$, was calculated to be 80%, corresponding to a THz power modulation of 96%. Additionally, we measured the THz amplitude modulations under the orthogonal polarized pumps, using the same pump intensity of 40 GW/cm$^2$, as shown in Figure 3d. The results show that the modulation efficiency for the *p*-polarized (red) pump was approximately three times higher than that for the *s*-polarized (blue) pump. This increased efficiency is attributed to the effective excitation of the Brewster mode in ENZ film under *p*-polarized pump at an angle of 67°.

According to the time-varying nonlinear model, the transient modulation with THz rate introduces an additional channel for energy transfer between the generated THz waves and the time-varying ITO film. Consequently, linear frequency shifts in the nonlinear THz generation are anticipated when the probe pulse interacts with the ITO film at the rising and falling edge of its nonlinear susceptibility. To validate this hypothesis, we measured the THz electric fields at various pump-probe time delays with a 62-fs interval, while maintaining the sampling range, pump intensity and polarization as in Fig. 3a. The THz temporal waveforms and their corresponding spectra are displayed in Figure 4a and Figure 4b, respectively. As the time delay increases, the frequencies of the THz peak amplitudes alternate between approximately 1.0 THz (indicated by the pink area) and around 1.37 THz (the blue area). The normalized THz spectra plotted in Figure 4c clearly illustrate a frequency blue shift, confirming that the generated THz photon energy acquire an ultrafast redistribution in response to the temporal modulation. Figure 4d presents a quantitative analysis of the central THz frequency as a function of time delay, where the central frequency is defined as the mathematical expectation of the THz spectrum:

$$v_c = \frac{\int_{v_1}^{v_2} v |A(v)|^2 dv}{\int_{v_1}^{v_2} |A(v)|^2 dv}, \tag{7}$$

with $v_1 = 0.1$ THz, $v_2 = 3.0$ THz defining the integral range due to the limited dynamic range of the experimental setup. We observed a significant dependence of the central frequency on the time delay. Notably, two peaks of frequency blue shift occur around 0.5 ps and 0.8 ps, corresponding to the rapid suppression and recovery, as shown in Fig. 3a. Intriguingly, a frequency shift valley was observed around 0.69 ps, where the maximal THz amplitude modulation was achieved, corresponding to the minimal modulation rate. These features mentioned above support our intuitive interpretation of the underlying physics governing the linear frequency shift in a nonlinear interaction of the time-varying media. Specifically, the ultrafast temporal modulation on the nonlinear susceptibility relaxes the constrains of the law of conservation of energy, and establishes a channel for the energy transferring through creating a temporal boundary. This can be regarded as a generalization of those processes in the linear optical regime. As a result, the relative frequency shift of 15% is achieved with the definition of $M_\nu = \frac{\nu_{modu} - \nu_{static}}{\nu_{static}} \times 100\%$, where $\nu_{static}$ and $\nu_{modu}$ represent the THz central frequency without pump (orange dashed line) and the largest THz central frequency with pump intensity of 40 GW/cm² (blue solid line), respectively.

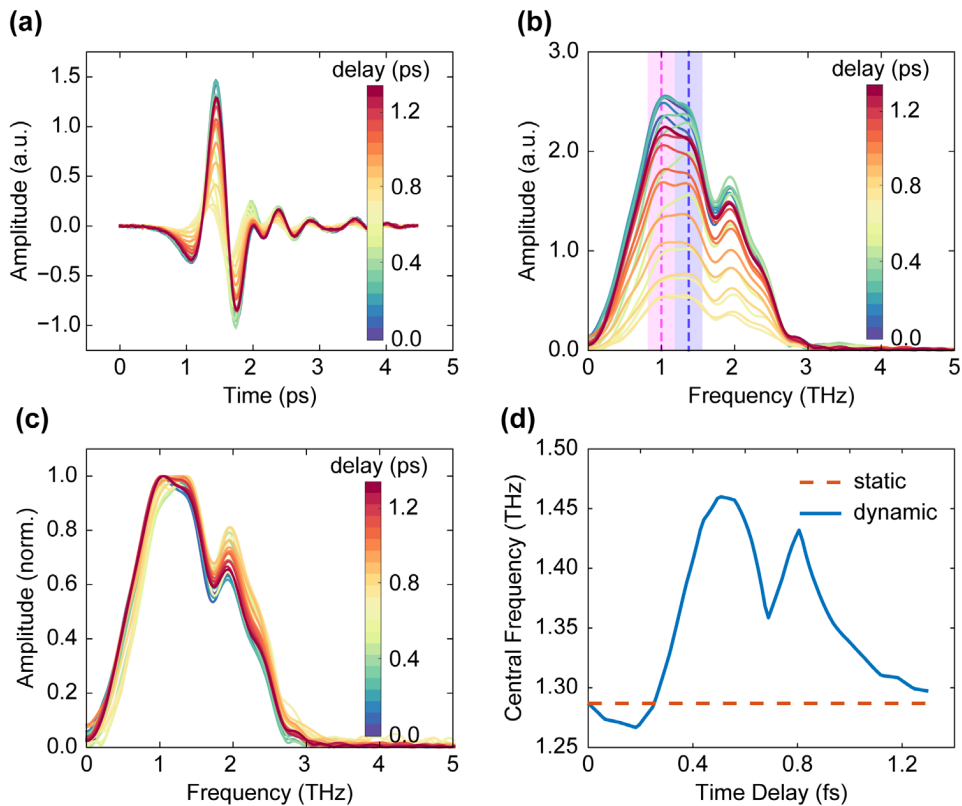

**Figure 4.** Measured THz central frequency shifts versus time delay under the pump intensity of 40 GW/cm². (a) THz electric field traces measured under various time delay between the pump pulses and the probe pulses, where the time delay was denoted by the color bar. (b) Corresponding THz spectra of temporal signals in Fig. 4(a). The spectra modulation with time delay was observed, where the spectral peaks evolve from the pink region to the purple region as the time delay increases, and then return to the pink region. (c) The normalized THz spectra were plotted and distinct frequency shifts as time delay increases were observed. (d) The static central frequency (orange dashed line) without pump and the dynamical central frequency (blue solid line) under pump intensity of 40 GW/cm². Two maxima at 0.5 ps and 0.8 ps sandwiching a minimum at 0.69 ps of the central frequency shift were observed, relating to the maxima and minimum of the amplitude modulation rate in Fig. 3(a), respectively.

The numerical simulations based on the dynamical nonlinear polarization model qualitatively reproduce the experimental results of amplitude and frequency modulations. In these simulations, we comparatively studied the roles of various linear susceptibilities that contribute to the nonlinear susceptibility. It was found that the time-varying linear susceptibility in NIR band primarily governs the THz amplitude and frequency modulations, while the contribution of the susceptibility in the THz band is negligible (see supplementary materials note 1).

In addition to modulating THz amplitude and frequency, we also experimentally demonstrated the ability to manipulate the polarization through time-varying nonlinearity. Based on our previous research, the polarization of the generated THz waves from the homogeneous and isotropic ITO film is determined by two independent tensor elements of $\chi^{(2)}$: $\chi_1 = \chi_{zzz}$, $\chi_2 = \chi_{xzx} = \chi_{xxz} = \chi_{yzy} = \chi_{yyz}$. In the aforementioned experiments, only the tensor elements associated with the $x$ and $z$ coordinates were used, with oblique incidence of *p*-polarized probe pulses, resulting in the generation of only *p*-polarized THz waves. In order to acquire the degree of freedom of polarization manipulation, the circularly polarized (CP) probe pulses were used to access the *y*-coordinate dependent tensor elements. Assuming a relationship between $\chi_1$ and $\chi_2$ given by $\chi_1 = \eta e^{i\zeta}\chi_2$, the ratio of the *y*-component and the *x*-component of the THz electric field can be expressed as:

$$\frac{E_y}{E_x} = \sigma \frac{A_y}{A_x} e^{i\delta}, \tag{8}$$

where $\sigma = \pm 1$ represent the CP handedness of probe pulses, with + 1 and – 1 denoting left-handed CP (LCP) and right-handed CP (RCP). The amplitude ratio $A_y / A_x$ and phase difference $\delta$ are functions of $\eta$ and $\zeta$ (see supplementary materials note 4). Consequently, a time-varying relative complex amplitude of $\chi_1$ and $\chi_2$ can enable the temporal manipulated THz polarization.

The measured polarization state of the THz electric field is represented using Stokes parameters[37], which are related with the Jones vectors parameters by:

$$\begin{aligned} S_0 &= A_x^2 + A_y^2 \\ S_1 &= \left(A_x^2 - A_y^2\right) \\ S_2 &= 2\sigma A_x A_y \cos(\delta) \\ S_3 &= 2\sigma A_x A_y \sin(\delta) \end{aligned}. \tag{9}$$

As shown in Figure 5(a), the Stokes parameters, normalized by $S_0$, were plotted on the Poincaré sphere within the 1.0 – 1.6 THz range, using the coordinate system of $S_1S_2S_3$. The corresponding THz frequencies are indicated by the color bar on the right column. As the time delay increases, shown by the black arrows, the Stokes parameters of the THz waves induced by the LCP (long dashed line) and the RCP (short dashed line) probe pulses exhibit antisymmetric evolution paths on the backside of the Poincaré sphere. This behavior is consistent with the prediction of equation (9), which indicates that the signs of $S_2$ and $S_3$ will reverse when the handedness of the probe pulses is opposite, while $S_1$ remains unaffected. However, under *p*-polarized probe excitation, the measured data deviates from the expected point of [1,0,0], likely due to measurement errors caused by system noise and imperfect p-polarization of the probe pulses.

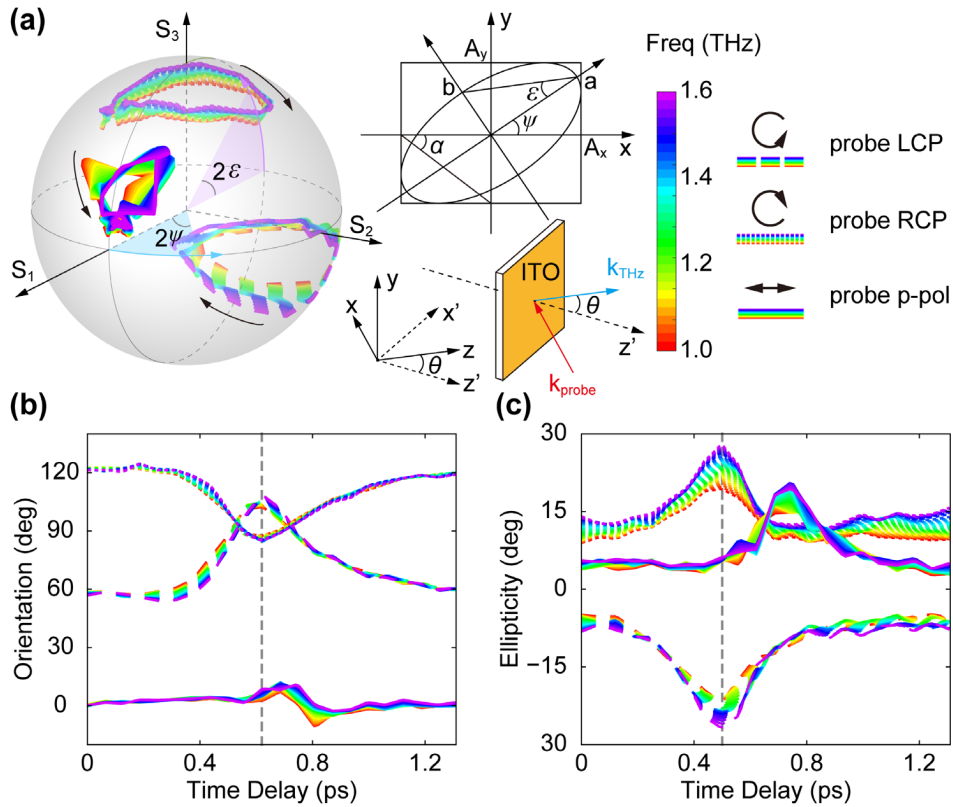

**Figure 5.** Measured THz polarization modulations versus time delay under the pump intensity of 40 GW/cm$^2$. (a) Broadband THz polarization state evolutions, excited by LCP (long dashed lines), RCP (short dashed lines) and p-polarized (solid lines) probe pulses, are depicted by normalized Stokes parameters on a Poincaré sphere. The THz frequency ranging from 1.0 to 1.6 THz is represented by the color bar. The polarization ellipse in the upper panel of the middle column illustrates the definitions of the orientation angle $\psi$ and ellipticity $\varepsilon$ on the Poincaré sphere, while the lower panel shows a schematic of the coordinate definition used in the discussion of polarization modulation. (b) The orientation angle versus time delay extracted from the Poincaré sphere. (c) The ellipticity versus time delay extracted from the Poincaré sphere.

The orientation angle $\psi$ and ellipticity $\varepsilon$ of the THz polarization ellipse were directly extracted from the Poincaré sphere and plotted in figure 5(b) and figure 5(c), respectively. The $\psi$ and $\varepsilon$ were defined by the corresponding geometric angles on the Poincaré sphere and the polarization ellipse, where the polarization ellipse was described in the plane facing to the opposite direction of the THz wave vector. The traces of the THz orientation angle and ellipticity versus the time delay under a specified probe polarization shows two peaks around 0.62 ps and 0.5 ps indicated by the gray dashed lines, respectively. This observation suggests a complex temporal modulation of the relative complex amplitude of $\chi_1$ and $\chi_2$. Notably, the ellipticity exhibits greater dispersion than the orientation angle, which is a natural outcome of the dispersive nature of equation (8) and the intrinsic constraint relationship between the amplitude ratio and phase difference (see supplementary materials note 5).

**Conclusion**

In summary, we have experimentally investigated the time-varying effects that occur within a single-cycle duration of nonlinear THz oscillation. We observed significant broadband THz amplitude modulation and central frequency shift, driven by the abrupt changes in the complex amplitude of $\chi^{(2)}$ during THz emission. Additionally, the relative complex amplitude of different tensor elements of $\chi^{(2)}$ further enabled the temporal manipulation of the THz polarization, with the polarization evolution path governed by the handedness of the probe pulses. Our work advances the study of time-varying phenomena into the nonlinear regime, offering a novel strategy to achieve the single-cycle modulation through cross-scale nonlinear conversion.

**Acknowledgements**

This work is supported by the National Natural Science Foundation of China (Grant Nos. 62075158, 62025504, 61935015, 62405215 and 62135008); China Postdoctoral Science Foundation (2024M752359); Postdoctoral Fellowship Program of CPSF (GZC20241200); Yunnan Expert Workstation (Grant No. 202205AF150008).

**Competing Interests Statement**

The authors declare that they have no competing interests.

**Methods**

First, the ITO film was deposited onto a 0.7 mm-thick optical glass substrate using e-beam evaporation at a deposition rate of 3 Å/s and a high voltage of 5 kV. Next, the deposited 50 nm-thick ITO film was placed in a tube furnace. The ITO film was preheated from room temperature to 100 °C over 5 minutes, followed by rapid heating to 500°C within 10 seconds. After maintaining this high temperature for 2 hours, the ITO film was cooled back to room temperature over 5 hours. This annealing process significantly increased the carrier concentration in the ITO film, blue shifting its ENZ wavelength to 1563 nm, as shown in Figure 2(a).

# References

(1) Yuan, L.; Lin, Q.; Xiao, M.; Fan, S. Synthetic dimension in photonics. *Optica* **2018**, *5*, 1396-1405.
(2) Emanuele, G.; Romain, T.; Shixiong, Y.; Huanan, L.; Stefano, V.; Paloma, A. H.; et al. Photonics of time-varying media. *Adv. Photonics* **2022**, *4*, 014002.
(3) Engheta, N. Four-dimensional optics using time-varying metamaterials. *Science* **2023**, *379*, 1190-1191.
(4) Ortega-Gomez, A.; Lobet, M.; Vázquez-Lozano, J. E.; Liberal, I. Tutorial on the conservation of momentum in photonic time-varying media [Invited]. *Opt. Mater. Express* **2023**, *13*, 1598-1608.
(5) Mendonça, J. T.; Shukla, P. K. Time Refraction and Time Reflection: Two Basic Concepts. *Physica Scripta* **2002**, *65*, 160.
(6) Xiao, Y.; Maywar, D. N.; Agrawal, G. P. Reflection and transmission of electromagnetic waves at a temporal boundary. *Opt. Lett.* **2014**, *39*, 574-577.
(7) Khurgin, J. B.; Clerici, M.; Bruno, V.; Caspani, L.; DeVault, C.; Kim, J.; et al. Adiabatic frequency shifting in epsilon-near-zero materials: the role of group velocity. *Optica* **2020**, *7*, 226-231.
(8) Zhou, Y.; Alam, M. Z.; Karimi, M.; Upham, J.; Reshef, O.; Liu, C.; et al. Broadband frequency translation through time refraction in an epsilon-near-zero material. *Nat. Commun.* **2020**, *11*, 2180.
(9) Bohn, J.; Luk, T. S.; Horsley, S.; Hendry, E. Spatiotemporal refraction of light in an epsilon-near-zero indium tin oxide layer: frequency shifting effects arising from interfaces. *Optica* **2021**, *8*, 1532-1537.
(10) Tirole, R.; Vezzoli, S.; Galiffi, E.; Robertson, I.; Maurice, D.; Tilmann, B.; et al. Double-slit time diffraction at optical frequencies. *Nat. Phys.* **2023**, *19*, 999-1002.
(11) Reyes-Ayona, J. R.; Halevi, P. Observation of genuine wave vector (k or β) gap in a dynamic transmission line and temporal photonic crystals. *Appl. Phys. Lett.* **2015**, *107*, 074101.
(12) Lustig, E.; Sharabi, Y.; Segev, M. Topological aspects of photonic time crystals. *Optica* **2018**, *5*, 1390-1395.
(13) Sharabi, Y.; Dikopoltsev, A.; Lustig, E.; Lumer, Y.; Segev, M. Spatiotemporal photonic crystals. *Optica* **2022**, *9*, 585-592.
(14) Zurita-Sánchez, J. R.; Halevi, P.; Cervantes-González, J. C. Reflection and transmission of a wave incident on a slab with a time-periodic dielectric function $\epsilon(t)$. *Physical Review A* **2009**, *79*, 053821.
(15) Lyubarov, M.; Lumer, Y.; Dikopoltsev, A.; Lustig, E.; Sharabi, Y.; Segev, M. Amplified emission and lasing in photonic time crystals. *Science* **2022**, *377*, 425-428.
(16) Wang, X.; Mirmoosa, M. S.; Asadchy, V. S.; Rockstuhl, C.; Fan, S.; Tretyakov, S. A. Metasurface-based realization of photonic time crystals. *Science Advances* **2023**, *9*, eadg7541.
(17) Shaltout, A.; Kildishev, A.; Shalaev, V. Time-varying metasurfaces and Lorentz non-reciprocity. *Opt. Mater. Express* **2015**, *5*, 2459-2467.
(18) Sounas, D. L.; Alù, A. Non-reciprocal photonics based on time modulation. *Nat. Photonics* **2017**, *11*, 774-783.
(19) Koutserimpas, T. T.; Fleury, R. Nonreciprocal Gain in Non-Hermitian Time-Floquet Systems. *Phys. Rev. Lett.* **2018**, *120*, 087401.
(20) Bacot, V.; Labousse, M.; Eddi, A.; Fink, M.; Fort, E. Time reversal and holography with spacetime transformations. *Nat. Phys.* **2016**, *12*, 972-977.
(21) Pacheco-Peña, V.; Engheta, N. Temporal aiming. *Light: Sci. Appl.* **2020**, *9*, 129.
(22) Akbarzadeh, A.; Chamanara, N.; Caloz, C. Inverse prism based on temporal discontinuity and spatial dispersion. *Opt. Lett.* **2018**, *43*, 3297-3300.
(23) Reshef, O.; De Leon, I.; Alam, M. Z.; Boyd, R. W. Nonlinear optical effects in epsilon-near-zero media. *Nat. Rev. Phys.* **2019**, *4*, 535-551.
(24) Jaffray, W.; Saha, S.; Shalaev, V. M.; Boltasseva, A.; Ferrera, M. Transparent conducting oxides: from all-dielectric plasmonics to a new paradigm in integrated photonics. *Adv. Opt. Photonics* **2022**, *14*, 148-208.

(25) Alam, M. Z.; De Leon, I.; Boyd, R. W. Large optical nonlinearity of indium tin oxide in its epsilon-near-zero region. *Science* **2016**, *352*, 795-797.

(26) Alam, M. Z.; Schulz, S. A.; Upham, J.; De Leon, I.; Boyd, R. W. Large optical nonlinearity of nanoantennas coupled to an epsilon-near-zero material. *Nat. Photonics* **2018**, *12*, 79-83.

(27) Tirole, R.; Vezzoli, S.; Saxena, D.; Yang, S.; Raziman, T. V.; Galiffi, E.; et al. Second harmonic generation at a time-varying interface. *Nat. Commun.* **2024**, *15*, 7752.

(28) Pang, K.; Alam, M. Z.; Zhou, Y.; Liu, C.; Reshef, O.; Manukyan, K.; et al. Adiabatic Frequency Conversion Using a Time-Varying Epsilon-Near-Zero Metasurface. *Nano Lett.* **2021**, *21*, 5907-5913.

(29) Lee, K.; Son, J.; Park, J.; Kang, B.; Jeon, W.; Rotermund, F.; Min, B. Linear frequency conversion via sudden merging of meta-atoms in time-variant metasurfaces. *Nat. Photonics* **2018**, *12*, 765-773.

(30) Cong, L.; Han, J.; Zhang, W.; Singh, R. Temporal loss boundary engineered photonic cavity. *Nat. Commun.* **2021**, *12*, 6940.

(31) Xu, G.; Xing, H.; Lu, D.; Fan, J.; Xue, Z.; Shum, P. P.; Cong, L. Linear Terahertz Frequency Conversion in a Temporal-Boundary Metasurface. *Laser Photonics Rev.* **2024**, *18*, 2301294.

(32) Won, R. It's a matter of time. *Nat. Photonics* **2023**, *17*, 209-210.

(33) Boyd, R. W. *Nonlinear optics*; Academic press, 2020.

(34) Jia, W.; Liu, M.; Lu, Y.; Feng, X.; Wang, Q.; Zhang, X.; et al. Broadband terahertz wave generation from an epsilon-near-zero material. *Light: Sci. Appl.* **2021**, *10*, 11.

(35) Taliercio, T.; Guilengui, V. N.; Cerutti, L.; Tournié, E.; Greffet, J.-J. Brewster "mode" in highly doped semiconductor layers: an all-optical technique to monitor doping concentration. *Opt. Express* **2014**, *22*, 24294-24303.

(36) Guo, P.; Schaller, R. D.; Ketterson, J. B.; Chang, R. P. H. Ultrafast switching of tunable infrared plasmons in indium tin oxide nanorod arrays with large absolute amplitude. *Nat. Photonics* **2016**, *10*, 267-273.

(37) Born, M.; Wolf, E. *Principles of optics: electromagnetic theory of propagation, interference and diffraction of light*; Elsevier, 2013.

# Supplementary Materials for

# Time-varying Nonlinear Effects in Terahertz Generation

Yongchang Lu[1], Xueqian Zhang[1]*, Haidi Qiu[1], Li Niu[1], Xieyu Chen[1], Quan Xu[1], Weili Zhang[2], Shuang Zhang[3]*, Jiaguang Han[1]*

[1]Center for Terahertz waves and College of Precision Instrument and Optoelectronics Engineering, Tianjin University and the Key Laboratory of Optoelectronics Information and Technology (Ministry of Education), Tianjin 300072, China.

[2]School of Electrical and Computer Engineering, Oklahoma State University, Stillwater, Oklahoma 74078, USA.

[3]Department of Physics, Faculty of Science, University of Hong Kong, Hong Kang

*Corresponding authors. Email: alearn1988@tju.edu.cn (XQ. Zhang), shuzhang@hku.hk (S. Zhang), jiaghan@tju.edu.cn (J. Han)

## Note 1. Numerical simulations for THz generation in time-varying media

The relative permittivity of the Drude type time-varying media is described by

$$\varepsilon_r(t-t') = \varepsilon_\infty - \frac{\omega_{p,0}^2\left[1-\Delta\omega_p(t-t')\right]^2}{\omega(\omega+i\gamma)}, \quad (S1)$$

where $\varepsilon_\infty = 2.7958$, $\omega_{p,0} = 2.0455 \times 10^{15}$ rad/s, $\gamma = 2.1240 \times 10^{14}$ rad/s are the infinite frequency permittivity, the static plasma frequency and the collision frequency. The time-varying effects are introduced by the term of $\Delta\omega_p(t-t')$, which represents the relative change of the plasma frequency induced by the optical pumping at $t'$. To simplify the numerical calculations, the $\Delta\omega_p(t-t')$ is phenomenologically modeled by a response function as

$$\Delta\omega_p(t-t') = \frac{\rho}{\left[1+e^{-(t-t')/\tau_1}\right]\left[1+e^{(t-t')/\tau_2}\right]} \quad (S2)$$

where $\tau_1 = 0.07$ ps and $\tau_2 = 0.25$ ps are two empirical parameters describing the thermalization time and the relaxation time of the carriers in ITO film, and $\rho$ represents the change in plasma frequency. In this simulation, $\rho$ is estimated to 0.2158, resulting in a maximal $\Delta\omega_p(t-t')$ change of 7%, with which the experimental results are well reproduced. Based on the time-varying plasma frequency, as well as the equation (5) and equation (6) in the main text, the spectra of the generated THz waves are calculated.

As shown in figure S1(a), the amplitudes of the THz spectra undergo significant modulation as the pump-probe time delay $t'$ changes. The normalized spectra are plotted in figure S1(b), where the frequency shifts are observed within the bandwidth of the generated THz waves. To determine the factors responsible for the broadband THz amplitude and frequency modulations, the dependences of three linear susceptibilities forming the second-order nonlinear susceptibility on the probe time $t$ and pump-probe time delay $t'$ are separately considering. Based on case 1, where the $t$-dependence and $t'$-dependence of the linear susceptibilities in both the NIR and THz bands are considered (as shown in fig. S1(a) and fig. S1(b)), the *t*-dependence of the NIR susceptibility is omitted in case 2, leading to the calculated THz spectra shown in fig. S1(c) and fig. S1(d). It can be seen that the amplitude modulations retain largely intact, while the frequency shifts are nearly negligible. In case 3, where the *t'*-dependence of NIR susceptibility is further removed, only a small amplitude modulation retains, as shown in figure S1(e) and figure S1(f). From the aforementioned three cases, it can be inferred that the *t*-dependence and *t'*-dependence of NIR susceptibility dominate the frequency shifts and the amplitude modulations of the generated THz waves, respectively. To intuitively display the variation of the THz amplitude and frequency with the pump-probe time delay, the peak amplitudes and central frequencies of the THz spectra are calculated and illustrated in figure S1(g) and figure S1(h), respectively. Here, the central frequency is defined as the mathematical expectation of the THz spectrum as

$$v_c = \frac{\int_0^{+\infty} v |E(v)|^2 dv}{\int_0^{+\infty} |E(v)|^2 dv}. \tag{S3}$$

Clearly, the amplitude modulations undergo asymmetric suppression and recovery dynamics as the time delay increases, where the maximal modulations are observed around 0.1 ps in all of the three cases. The hysteresis of the maximal amplitude modulation relative to the zero delay is due to the non-instantaneous response of the material, while the asymmetric feature is attributed to the difference of the excitation and relaxation processes of the carriers, which are described by the two parameters of $\tau_1$ and $\tau_2$. The central frequency in case 1 not only has the largest shift, but also exhibits a significant frequency shift valley around 0.06 ps. This complex frequency modulation qualitatively reproduces the experimental results, confirming that the frequency shifts in time-varying nonlinear process mainly depend on the time-varying linear susceptibility in NIR band. Furthermore, the changes of the modulated THz amplitudes and central frequencies compared to that of unmodulated case are recorded in figure S1(i), where nine cases are summarized in total, as shown in the inset table, definitely highlighting the major role of the *t*-dependence of NIR susceptibility in frequency modulation.

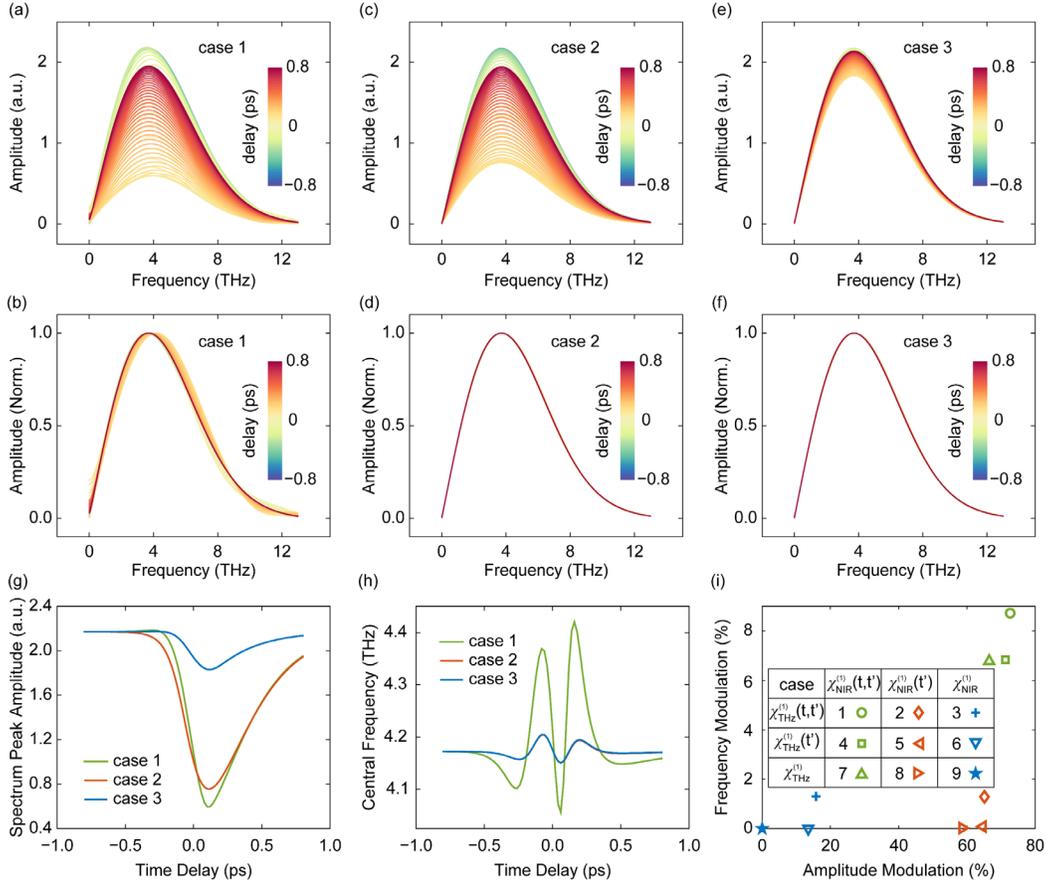

Figure S1. Numerical simulations of the modulated THz spectra versus the pump-probe time delay under various time-varying conditions. The simulated THz spectra in three cases are shown in figures S1(a), (c) and (e), where the corresponding simulation conditions are indicated by the symbols of green circle, orange diamond and blue cross in figure S1(i), respectively. Each row in the table of figure S1(i) represents the same *t*- or *t′*-dependences of linear susceptibility in the THz band, while each column represents the same *t*- or *t′*-dependences of linear susceptibility in the NIR band. Figures S1(b), (d) and (f) show the normalized THz spectra of that in figures S1(a), (c) and (e). Figure S1(g) illustrates the variation of THz peak amplitudes of three cases with the time delay, where the major role of the *t′*-dependence of NIR susceptibility for the THz amplitude modulation

is recognized. Figure S1(h) illustrated the central frequencies of these THz spectra, where the $t$-dependence of NIR susceptibility for the THz frequency modulation is recognized. Figure S1(i) further summarizes the relative amplitude and central frequency shifts for the other six cases of various $t$- or $t'$-dependences of the linear susceptibility.

**Note 2. Excitation of Brewster mode in ITO film**

To efficiently modulate the optical property of ITO film in NIR band, we performed a numerical simulation to optimize the incident angle of the pump pulse. As shown in figure S2, the absorptance of the pump energy in ITO film is a function of the incident angle, where the peak absorptance occurs at the angle of 65°. In the experiment, the actual incident angle is about 67°, so that the Brewster mode is efficiently excited, reducing the requirement of the incident pump intensity.

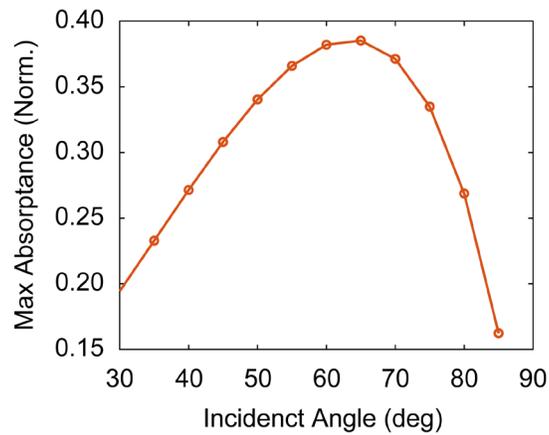

Figure S2. The absorptance of incident light in NIR band as a function of the incident angle. The peak absorptance around the incident angle of 65° indicates the efficient excitation of the Brewster mode in the ITO film, where the electric enhancement significantly facilitates the modulation of the optical properties of ITO film.

**Note 3. Optimization of the probe intensity**

The nonlinear interaction between the probe pulse and the ITO film becomes significant as the intensity of probe pulse increases. However, excessive probe intensity will affect the optical property of the ITO film, leading to difficulties to separate the influence of pump pulses in the temporal modulation of ITO film. Therefore, we conducted experiments to determine a moderate probe intensity to balance the excitation of nonlinear interactions and the modulation to the ITO film. Figure S3(a) shows the measured THz electric fields with probe intensities ranging from 4.7 to 18.7 GW/cm$^2$ with an increment of 1.17 GW/cm$^2$. The peak amplitudes are plotted in figure S3(b), where a linear relationship is observed with probe intensity ranging from 4.7 to 11.7 GW/cm$^2$, while a saturation behavior appears when the probe intensity exceed 11.7 GW/cm$^2$. In this situation, the probe intensity in the other experiments is determined to be 7 GW/cm$^2$.

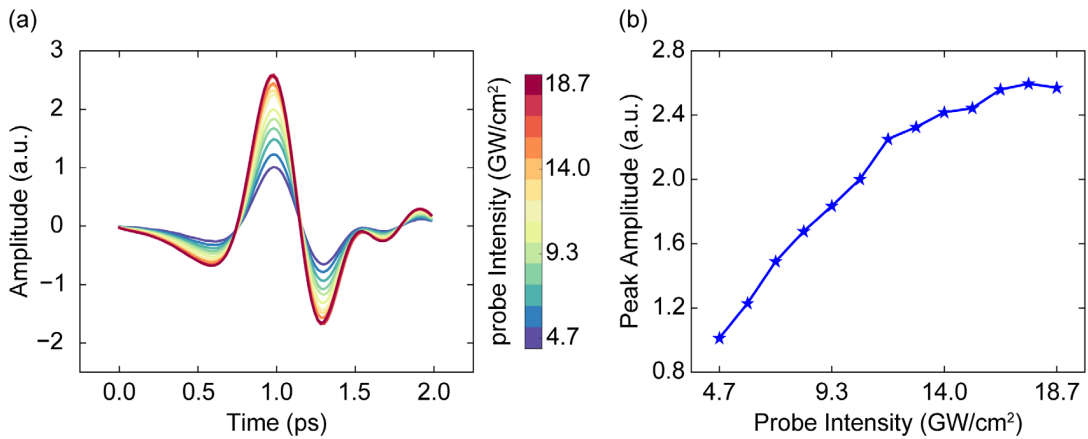

Figure S3. (a) Measured THz electric fields with probe intensity ranging from 4.7 to 18.7 GW/cm$^2$. (b) The peak amplitudes extracted from the temporal traces of the THz electric fields in figure S3(a), where a linear relationship is observed with probe intensity ranging from 4.7 to 11.7 GW/cm$^2$.

**Note 4. THz polarization manipulation based on the relative complex amplitude of the tensor elements of nonlinear susceptibility**

Two independent tensor elements of $\chi_1 = \chi_{z'z'z'}$ and $\chi_2 = \chi_{x'z'x'} = \chi_{x'x'z'} = \chi_{y'z'y'} = \chi_{y'y'z'}$ in ITO film play the major role in determining the polarization of the generated THz waves. Generally, these two elements are complex numbers, representing the dispersion and absorption characteristics of ITO in the frequency domain. Assuming a relationship between the two elements by:

$$\chi_1 = \eta e^{i\zeta} \chi_2, \tag{S4}$$

the polarization of the generated THz waves can be derived. With a circularly polarized probe pulse at a 45° incident angle, the complex amplitudes of the three electric field components in the ITO film can be expressed in the local coordinate system as:

$$\begin{aligned} E_{x'} &\propto \eta_{x'} \\ E_{y'} &\propto e^{i\xi} \\ E_{z'} &\propto -\eta_{z'} e^{i\beta} \end{aligned}, \tag{S5}$$

where $\eta_{x'}$ and $\eta_{z'}$ denote the electric field enhancement of the $x$ and $z$ components relative to the $y$ component. Here, $\xi = \sigma\pi/2$ indicates the handedness of the probe pulses, with $\sigma = +1$ representing left circularly-polarized (LCP) and $\sigma = -1$ representing right circularly-polarized (RCP), while $\beta$ is the phase difference between the $z$ and $x$ components. Therefore, the THz polarization components in the local coordinate system can be expressed as:

$$\begin{aligned} P_{x'} &= \chi_{x'z'x'} E_{z'} E_{x'}^* + \chi_{z'z'z'} E_{x'} E_{z'}^* = -2\eta_{x'}\eta_{z'}\chi_2 \cos(\beta) \\ P_{y'} &= \chi_{y'z'y'} E_{z'} E_{y'}^* + \chi_{y'y'z'} E_{y'} E_{z'}^* = -\eta_{z'}\chi_2 \cos(\beta - \xi). \\ P_{z'} &= \chi_{z'z'z'} E_{z'} E_{z'}^* = \eta \eta_{z'}^2 \chi_2 e^{i\zeta} \end{aligned} \tag{S6}$$

Through transforming the THz polarization into laboratory coordinate system, the ratio between the y and x components of THz electric field can be expressed as:

$$\frac{E_y}{E_x} = \frac{\sigma 2\sqrt{2} \sin(\beta)}{2\eta_x \cos(\beta) + \eta \eta_z e^{i\zeta}} \tag{S7}$$

The values of $\beta$, $\eta_x$ and $\eta_z$ can be obtained by numerical simulations, as such, the above equation

can be simplified as:

$$\frac{E_y}{E_x} = \sigma \frac{A_y}{A_x} e^{i\delta} \tag{S8}$$

where the amplitude ratio $A_y / A_x$ and the phase difference $\delta$ are functions of $\eta$ and $\zeta$, which is the equation (8) in the main text.

**Note 5. Dispersion behavior in the orientation and the ellipticity modulation of the THz waves**

According to the definitions of the Stokes parameters in the equation (9), as well as the geometry relationship in the polarization ellipse depicted in the middle column of figure 5(a) in the main text, we have the following equations respect to the orientation $\psi$ and the ellipticity $\varepsilon$:

$$\begin{aligned}\tan(2\psi) &= \tan[2\alpha(v)]\cos[\delta(v)] \\ \sin(2\varepsilon) &= \sin[2\alpha(v)]\sin[\delta(v)]\end{aligned} \quad (S9)$$

where the definition of $\tan(\alpha) = A_y / A_x$ is used, and $\delta$ is the phase difference between the $y$ and $x$ components of the generated THz waves. Here, $\alpha$ and $\delta$ are generally considered to be dispersive and expressed as functions of frequency $v$. Based on the experimental results shown in figure 5(b), the orientation of the THz polarization ellipse exhibits low dispersion, indicating that the dispersions of $\tan[2\alpha(v)]$ and $\cos[\delta(v)]$ largely compensate for each other. Due to $|\tan[2\alpha(v)]|$ and $|\sin[2\alpha(v)]|$ have the same monotonicity respect to $v$, while $|\cos[\delta(v)]|$ and $|\sin[\delta(v)]|$ have the opposite monotonicity, the dispersion of the product of $\sin[2\alpha(v)]$ and $\sin[\delta(v)]$ is inherently enhanced, resulting in the significant dispersion of the ellipticity as shown in figure 5(c).